\documentclass[amsmath,preprint,12pt]{elsarticle}

\usepackage{amssymb}
\usepackage[utf8]{inputenc}
\usepackage[T1]{fontenc} 
\usepackage{graphicx} 
\usepackage{bm} 
\usepackage{ulem}

\usepackage[hypertexnames=false,colorlinks=true,citecolor=blue,
			linkcolor=blue,urlcolor=blue]{hyperref}

\usepackage{xfrac}

\journal{Journal of Alloys and Compounds }

\begin{document}

\begin{frontmatter}



\title{Dirac Dispersions and Fermi Surface Nesting in LaCuSb$_{2}$}


\author[inst1,inst2]{Marcin~Rosmus\fnref{firstfoot}}
\author[inst1]{Natalia~Olszowska}
\author[inst3]{Zbigniew~Bukowski}
\author[inst4]{Przemys\l{}aw~Piekarz}
\author[inst4]{Andrzej~Ptok\fnref{secondfoot}}
\author[inst2]{Pawe\l{}~Starowicz\fnref{thirdfoot}}

\affiliation[inst1]{organization={Solaris National Synchrotron Radiation Centre, Jagiellonian University},
            addressline={Czerwone Maki 98}, 
            city={Kraków},
            postcode={30-392}, 
            country={Poland}}

\affiliation[inst2]{organization={Marian Smoluchowski Institute of Physics, Jagiellonian University},
            addressline={Prof. S. {\L}ojasiewicza 11}, 
            city={Kraków},
            postcode={PL-30348}, 
            country={Poland}}

\affiliation[inst3]{organization={Insitute of Low Temperature and Structure Research, Polish Academy of Sciences},
            addressline={P.O. Box 1410}, 
            city={Wroc\l{}aw},
            postcode={50-950}, 
            country={Poland}}

\affiliation[inst4]{organization={Institute of Nuclear Physics, Polish Academy of Sciences},
            addressline={W. E. Radzikowskiego 152}, 
            city={Kraków},
            postcode={PL-31342}, 
            country={Poland}}

\fntext[firstfoot]{marcin.rosmus@uj.edu.pl}
\fntext[secondfoot]{aptok@mmj.pl}
\fntext[thirdfoot]{pawel.starowicz@uj.edu.pl}

\begin{abstract}
LaCuSb$_{2}$ is a superconductor with a transition temperature of about $T_\text{c} = 0.9$~K and is a potential platform where Dirac fermions can be experimentally observed. In this paper, we report systematic high-resolution studies of its electronic structure using the angle-resolved photoemission spectroscopy (ARPES) technique supported by the DFT calculation. The Fermi surface consists of four branches, of which the two inner ones are more 3-dimensional and the theoretical calculations reproduce well the experiment. We observe several linear dispersions forming Dirac-like structures. The nodal lines are present in the system along ${\text{M}}$--${\text{A}}$ and ${\text{X}}$--${\text{R}}$ and Dirac crossings along ${\text{X}}$--${\text{R}}$ are observed by ARPES. Finally, the nesting between external Fermi surface pockets, which corresponds to charge density wave (CDW) modulation vector is enhanced in LaCuSb$_{2}$ as compared to LaAgSb$_{2}$, while CDW appears in the latter system.
\end{abstract}




\end{frontmatter}


\section{Introduction}
\label{sec:sample1}

A new chapter in solid-state physics was opened with the discovery of topological \mbox{insulators~\cite{1Hasan2010,2Haldane2017,3Bhardwaj2020}} 
 and the topic was subsequently expanded with nodal line semimetals~\cite{4Fang2016,5Lee2021,6Fang2015}, Dirac~\cite{7Armitage2018,8Liu2014,9Liu2014}, Weyl semimetals~\cite{10Lv2015,11Weng2015, 12Xu2015}, and other topologically non-trivial systems. The dispersion relation in the case of Dirac semimetals is described by the relativistic Dirac equation and manifests itself in the form of linear bands forming characteristic cones~\cite{13Burkov2016,14Burkov2011}. 
Such structures have been observed in Na$_{3}$Bi$_{8}$ and Cd$_{3}$As$_{2}$~\cite{9Liu2014}. 
A development of this idea is to extend the Dirac point in the second dimension so that it creates a nodal line \cite{15Yang2018}. 
Examples of systems in which nodal lines have been found are PtSnSb$_{4}$~\cite{16Wu2016}, (Zr/Hf)SiS~\cite{17Fu2019,18Chen2017}, InBi~\cite{19Ekahana2017}, AlBSb$_{2}$~\cite{20Takane2018}, and SrAsSb$_{3}$~\cite{21Song2020}.
Characteristic diamond-shaped Fermi surface (FS) is often observed in these materials~\cite{22Yen2021,23Nakamura2019,24Neupane2016,25bSabin2023}.

The attention of researchers was attracted by the $RT$Sb$_{2}$ ($R$: rare earth, $T$: the $d$-electron transition metal) family of intermetallic compounds with an interplay between superconductivity~\cite{25Lakshmi1996} and charge density waves (CDW)~\cite{27Song2003,28Singha2020,29Bosak2021}. 
Furthermore, the theoretical calculations suggest the appearance of Dirac-like bands in these compounds~\cite{30Ruszaa2018,31Hase2014}. Electronic structure of $RT$Sb$_{2}$ systems has already been studied by the angle-resolved photoemission spectroscopy (ARPES)\cite{32Shi2016,35Chamorro2019,33Rosmus2022, PhysRevB.108.245156, datta2023evidence}.
For LaAgSb$_{2}$, observation of Dirac-like structures and nesting of the FS has been reported~\cite{32Shi2016}, and more recently the nodal lines have been observed~\cite{33Rosmus2022}. 
A particularly interesting compound of this type is LaCuSb$_{2}$, which exhibits superconducting properties below $T_\text{c} = 0.9$~K~\cite{34Muro1997}.
Moreover, LaCuSb$_{2}$ is considered as another system with Dirac fermions as measurements of Shubnikov--de Haas quantum oscillations showed an electron effective mass of $0.065$~m$_{e}$ and transport measurements revealed linear magnetoresistance~\cite{35Chamorro2019}. 
In agreement with these predictions the first ARPES studies indicated that linear dispersion is present for certain bands~\cite{35Chamorro2019}.

In this paper, we present a detailed and systematic study of band structure and FS of LaCuSb$_{2}$ performed with ARPES technique and DFT calculations. 
Dispersions of Dirac fermions are found in new regions, which have not been identified previously~\cite{35Chamorro2019}. 
DFT calculations predicting the existence of nodal lines are in a good agreement with the experiment. 
The paper is organized as follows.
Details of the techniques used are provided in Sec~\ref{sec.method}.
Next, in Sec.~\ref{sec.res} we present and discuss our results.
Finally, a summary is provided in Sect.~\ref{sec.sum}.

\section{Methods and techniques}
\label{sec.method}

Single crystals of LaCuSb$_{2}$ were grown by the flux technique using Sb-rich antimony-copper melt as a flux. 
The excess of Cu was applied in order to prevent the formation of Cu-deficient LaCu$_{1-x}$Sb$_{2}$ phase. Lanthanum (purity $99.99$\%), copper (purity $99.99$\%) and antimony (purity $99.999$\%) were used as starting materials. 
The components were weighed in the atomic ratio La:Cu:Sb=1:1.5:13 and placed in an alumina crucible, which was then sealed in an evacuated silica tube. 
The ampoule was heated at $1100^{\circ}$C for $5$~h followed by slow cooling ($2-3^{\circ}$C/h) down to $700^{\circ}$C. 
At this temperature, the ampoule was flipped upside down in order to decant liquid Sb-Cu flux. 
Next, the ampoule was cooled to room temperature and the crucible was transferred to another silica tube, where the rest of Sb was removed from crystals by means of sublimation in a high dynamic vacuum at $600^{\circ}$C. 
Finally, single crystals were mechanically separated.

The obtained crystals were examined using a scanning electron microscope (SEM) Philips 515 and their chemical composition was determined with an energy dispersive X-ray (EDX) spectrometer PV9800. 
The phase purity and lattice parameters were determined from powder X-ray diffraction data obtained on pulverized single crystals.

High-resolution angle-resolved photoemission studies were performed at the URANOS beamline of Solaris Synchrotron~\cite{37Szlachetko2023}, Krak\'{o}w, Poland. 
The beamline is equipped with a Scienta-Omicron DA30-L electron analyzer. 
The geometry of the experimental setup is shown in Figure~\ref{f1}(c). 
The samples were cleaved in situ under ultrahigh vacuum at room temperature. The pressure during the ARPES measurement was below $5 \times 10^{-11}$~mbar and the temperature during the experiment was $14$~K.

The electronic structure was calculated using density functional theory (DFT) within the projector augmented-wave (PAW) method \cite{38Kresse1999} with the Vienna Ab initio Simulation Package ({\sc VASP})~\cite{38Kresse1999,39Kresse1996,40Kresse1994}. 
The exchange correlation potential was obtained by the generalized gradient approximation (GGA) in the Perdew--Burke--Enzerhof form~\cite{41Perdew1996}. 
Calculations were made using a $20\times20\times10$ Monkhorst–Pack {\bf k}-point mesh~\cite{42Monkhorst1976}. 
The kinetic energy cut-off for the plane-wave expansion was equal to $520$~eV. 
The energy convergence criteria used to relax the structures, with the conjugate gradient technique, were set at $10^{-8}$~eV and $10^{-6}$~eV for the electronic and ionic iterations, respectively.

The optimized structure was used to construct the tight-binding model in the maximally localized Wannier orbitals~\cite{marzari.vanderbilt.97,souza.marzari.01,marzari.mostofi.12}, using {\sc Wannier90} software~\cite{mostofi.jonathan.14,pizzi.vitale.20}.
In the discussed case, we started calculation from initial $p$-orbitals of Sb and $d$-orbitals of La and Cu, what corresponds to 32-orbitals tight binding model.
In order to study the surface states, the surface Green function for a semi-infinite system \cite{43LopezSancho1985} was calculated using {\sc WannierTools}~\cite{44Wu2018}.

\section{Results and discussion}
\label{sec.res}

\begin{figure*}
\centering
\includegraphics[width=\textwidth]{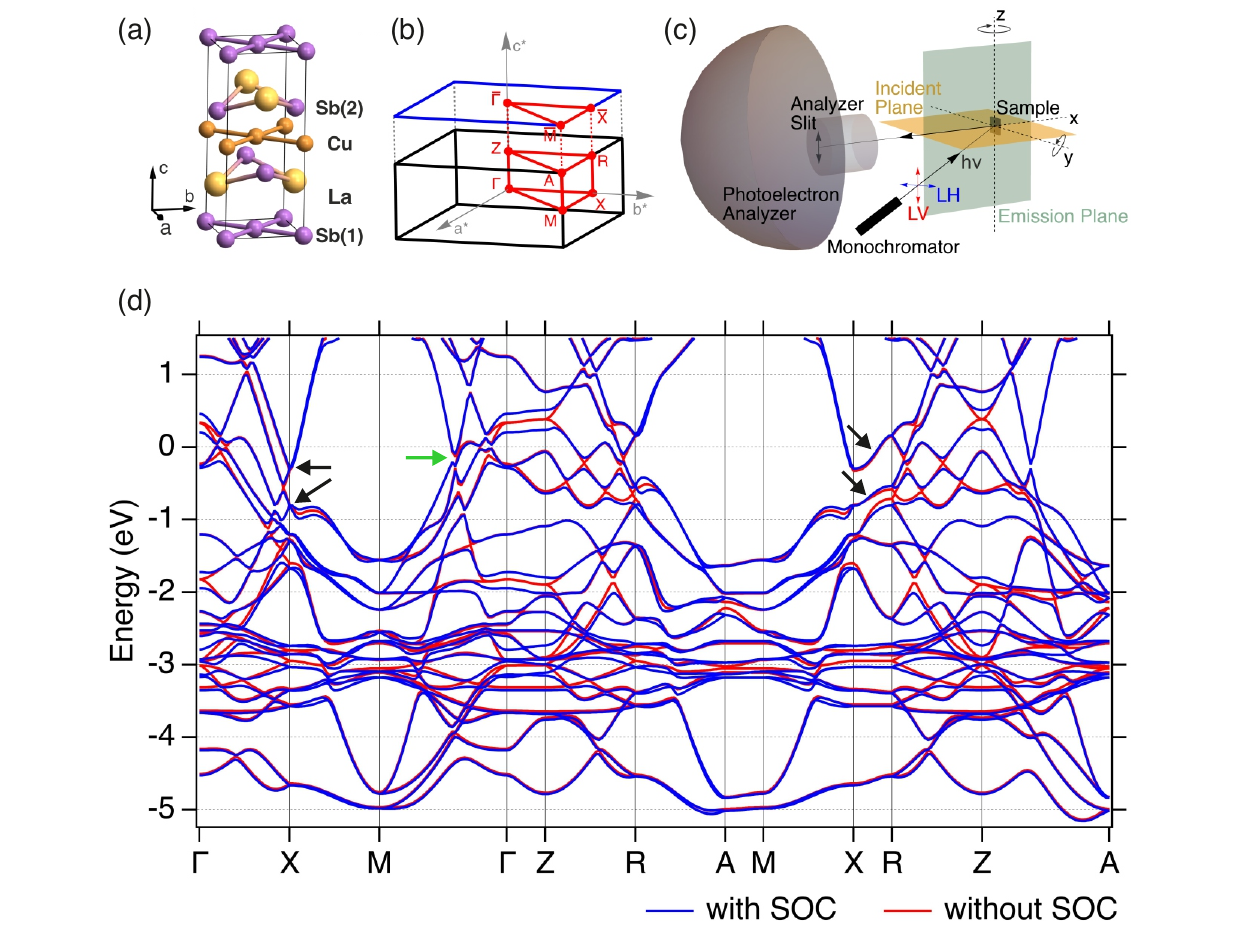}%
\caption{The tetragonal crystal structure of LaCuSb$_{2}$ (a) and the corresponding first Brillouin zone with their high-symmetry points (b).
The geometry of the ARPES measurement system (c). 
The bulk band structure (d) in the absence and presence of the spin--orbit coupling (blue and red lines, respectively). 
Black arrows indicate the position of the nodal line, while green arrow marks the location of the gapped structure.
}
\label{f1}
\end{figure*}

LaCuSb$_{2}$ crystallizes in the tetragonal, layered structure (P4/nmm, space group No. 129), consisting of Sb(2)-La layers related by glide symmetry and two-dimensional planes formed of either Cu or Sb(1) atoms.
Crystal structure of LaCuSb$_{2}$ is shown in Fig.~\ref{f1}(a). 
The lattice constants are equal to $a = 4.3828(2)$~\AA\ and $c = 10.2097(7)$~\AA~\cite{36Yang2005}. 
The tetragonal Brillouin zone corresponding to the lattice is shown in Fig.~\ref{f1}(b).

Bulk calculation of the band structure, shown in Fig.~\ref{f1}(d), with and without spin-orbit coupling (SOC) indicates that the electronic structure consists of several bands with linear dispersion along the $\Gamma$--${\text{X}}$ and $\Gamma$--${\text{M}}$ path. 
The existence of these bands may be associated with Dirac fermions. 
Moreover, several places can be distinguished where band crossing exists (e.g. in the proximity of the X point). 
As a consequence of the symmetry of the system, the nodal lines are realized in the ${\text{X}}$--${\text{R}}$ and ${\text{M}}$--${\text{A}}$ directions. 
At the X and R points, the band structure without SOC shows Dirac nodes that are part of the nodal line that crosses the Fermi energy [nodal lines are marked with black arrows in Fig.~\ref{f1}(d)]. 
The nodal line is also present after considering the spin-orbit interaction, no gap is observed in the structure in the ${\text{X}}$--${\text{R}}$ direction. 
The case is different in the direction of ${\Gamma}$--${\text{M}}$, in which band crossing is eliminated under the influence of SOC and a gapped structure is present [green arrow in Fig.~\ref{f1}(d)]. 

\begin{figure*}
\centering
\includegraphics[width=\textwidth]{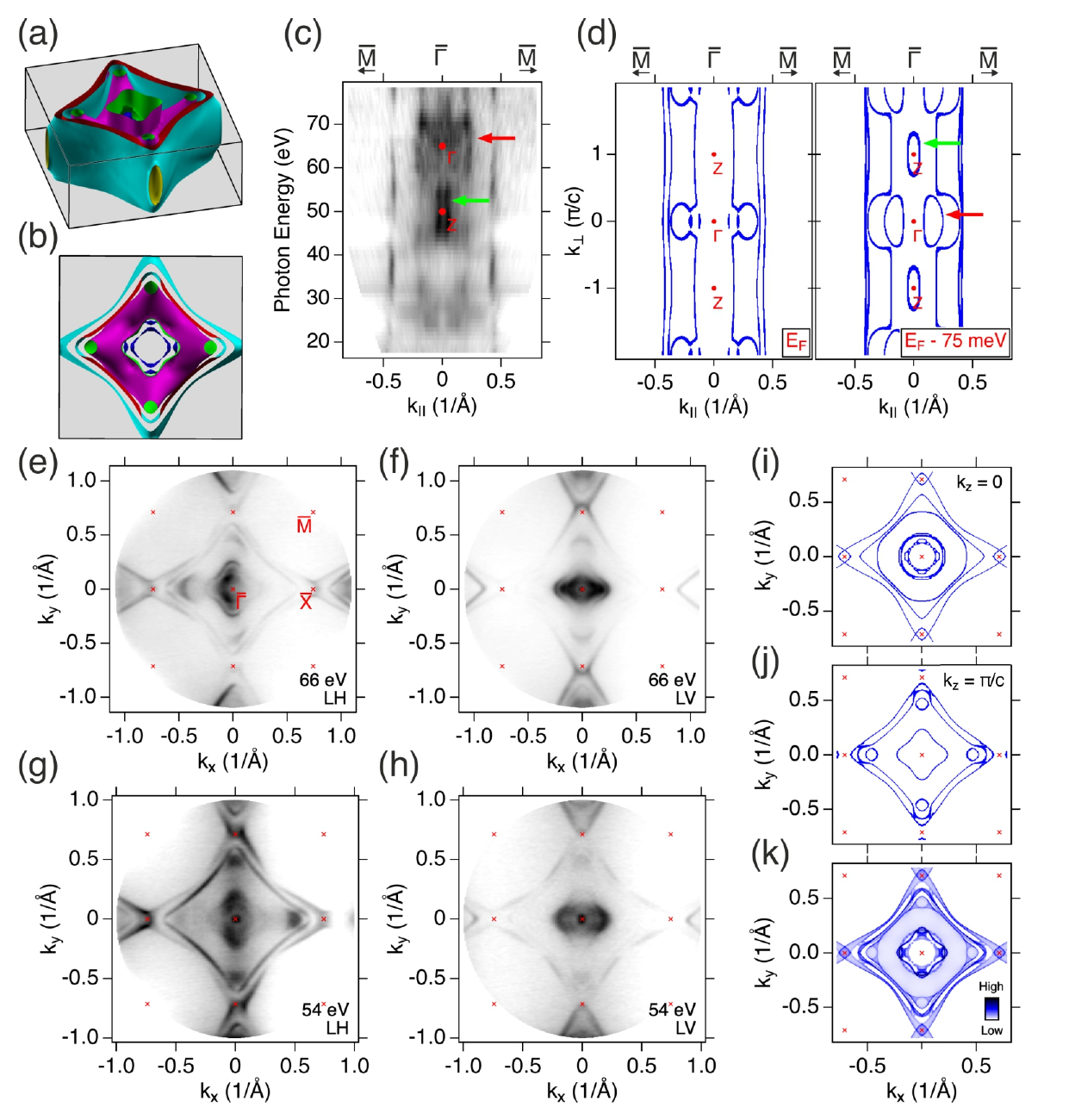}%
\caption{
Fermi Surface of LaCuSb$_{2}$. Three-dimensional view of the FS calculated without SOC (a), and its top view (b). 
Photon energy dependence map (c) along the $\bar{\text{M}}$--$\bar{\Gamma}$--$\bar{\text{M}}$ direction for the binding energy corresponding to the Fermi level. 
Calculated constant energy map (d) along the $\bar{\text{M}}$--$\bar{\Gamma}$--$\bar{\text{M}}$ direction.
Panels (e)-(h) present the FS obtained by ARPES for different photon energies and polarisations (LH – linear horizontal, LV – linear vertical). 
Crossection of the FS for $k_{z} = 0$ (i) and $k_{z} = \pi/c$ (j), while (k) presents projection of the bulk FS onto two-dimensional surface BZ. 
Red and green arrows in (c) and (d) indicate characteristic features of the experimental and theoretical spectrum.}
\label{f2}
\end{figure*}

According to the calculations, FS is composed of four bands, two of them forming the characteristic large diamond-like shape spanned approximately between X points, while the other two are strongly three-dimensional and centered at the $\Gamma$ point [Fig.~\ref{f2}(a) and (b)]. 
To fully understand the electronic structure of LaCuSb$_{2}$, we performed the photon energy dependence measurement to find the photon energy corresponding to the high-symmetry points.
At the collected FS, we can distinguish four bands, two of which have clear quasi-two-dimensional characters ($k_{\parallel} \sim \pm 0.4$~\AA$^{-1}$), and the other two create characteristic shapes, marked with the red and green arrow in Fig.~\ref{f2}(c). 
To compare ARPES results with the theoretical calculations we analyzed several constant energy maps from the vicinity of $E_{F}$. 
The map corresponding to the binding energy of $75$~meV revealed similar-looking structures, which allowed us to identify the location of the points of high symmetry. 
The difference of the $75$~meV between the measured and calculated map may be the effect of the energy resolution of the $k_{z}$ scan and the shifts of the simulated bands relative to the real ones. 
The calculations corresponding to $E_{F}$ and $E_{F} - 75$~meV are shown in Fig.~\ref{f2}(d), the green and red arrows correspond to analogous structures in the experimental map. 
The position of the $\Gamma$ point corresponds to photon energy about $66$~eV, and the Z point to about $50$~eV.
The experimental in-plane FSs are shown in Fig.~\ref{f2}(e)-(h). 
Measurements were performed using photon energies of 66 and $54$~eV with either linear vertical (LV) or linear horizontal (LH) polarization; details are specified in the figures. 
The intensity of the four pockets that form the FS strongly depends on the polarization of the light. 
The outer pockets are visible as two arc-shaped lines connecting the $\bar{\text{X}}$ points. 
The separation between them is clearly visible in each of the measurements. 
The inner pockets form an intricate structure centered around the $\bar\Gamma$ point. 
Comparison with the theoretical calculations corresponding to the planes including the points of high symmetry shows similarity for $k_{z} = 0$ and a large discrepancy for $k_{z} = \pi/c$ [Fig.~\ref{f2}(i) and (j)]. 
However, theoretical FS map integrated over all k$_{z}$ [Fig.~\ref{f2}(k)] shows much greater similarity to the experimental measurements, indicating a strong mixing of k$_{z}$ what has already been reported for similar compounds~\cite{17Fu2019,33Rosmus2022}.

\begin{figure*}
\centering
\includegraphics{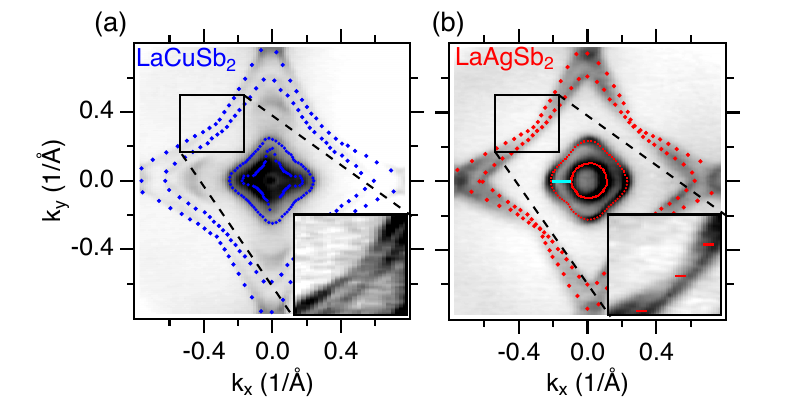}%
\caption{Comparison of Fermi surface nesting in (a) LaCuSb$_{2}$ and (b) LaAgSb$_{2}$. 
The red lines in (b) correspond to the length of the CDW vector $q_{1} = 0.026\cdot(2\pi/a)$, and the cyan line shows the possible nesting vector.}
\label{f3}
\end{figure*}

A nesting of the FS can play a role in the formation of CDW or superconducting ground state. 
Thus, it is instructive to study the nesting in superconducting LaCuSb$_{2}$ and compare it with the relative system LaAgSb$_{2}$, which exhibits CDW. 
A detailed description of the electronic structure of LaAgSb$_{2}$ was published by Rosmus {\it et al.} in Ref.~\cite{33Rosmus2022}. 
The main idea behind the study of nesting in the compound with Ag was related to the scenario of the Peierls transition. 
According to this concept, one should observe the nesting in a compound exhibiting CDW and its decay would be expected when CDW is weakened and superconductivity appears in the phase diagram. 
Nesting itself has already been reported for LaAgSb$_{2}$ by Shi {\it et al.} in Ref.~\cite{32Shi2016}.
They suggested that the nesting takes place between the bands forming the diamond-like pockets connecting X points in the FS, and the nesting vector is consistent with the CDW vector. 
A comparison of the FSs of LaCuSb$_{2}$ and LaAgSb$_{2}$ is shown in Fig.~\ref{f3}. 
These results were collected at $66$~eV (semiplanar cut crossing our proposed $\Gamma$ point) and obtained by adding LV and LH polarization data in order to get more symmetrical maps. 
The insets are focused on the region in which nesting may appear. 
Our high-resolution ARPES measurements indicate that this nesting is not perfect, and the shape of the considered pocket cannot be reproduced by a translation alone. 
In fact, only single points of the FS can be connected by the nesting vector, $q_{1}\sim 0.026 \cdot (2\pi/a)$ related to CDW modulation ~\cite{27Song2003} [red lines in Fig.\ref{f3}(b) - inset]. 
Moreover, we made attempts to measure the energy gap for LaAgSb$_{2}$ associated with the CDW. It was not observed in any of the bands (not shown) by ARPES measurements carried out at $T = 18$~K with energy resolution of 10 meV. In the case of internal pockets centered around the $\bar\Gamma$ point, nesting features can be found on their rounded  edges $q \sim0.089 \cdot (2\pi/a)$ [marked with cyan line in Fig.~\ref{f3}(b)]. The shape of the smallest pocket is significantly different in the case of the compound with Cu, where a removal of nesting is observed. 
However, the nesting vector found here between the internal pockets is  more than three times larger than that corresponding to CDW modulation measurements q$_{1}$, which is inconsistent with the Peierls scenario. 
Surprisingly, nesting at diamond-like pockets appears to be more perfect in the case of superconducting LaCuSb$_{2}$. 
In the case of this compound, we can speak of the similarity of diamond-like pockets sufficient to reconstruct them by shifting the appropriate FS fragments. That is a surprising result that contradicts simple expectations of a Peierls transition scenario in LaAgSb$_{2}$.

\begin{figure*}
\centering
\includegraphics[width=\textwidth]{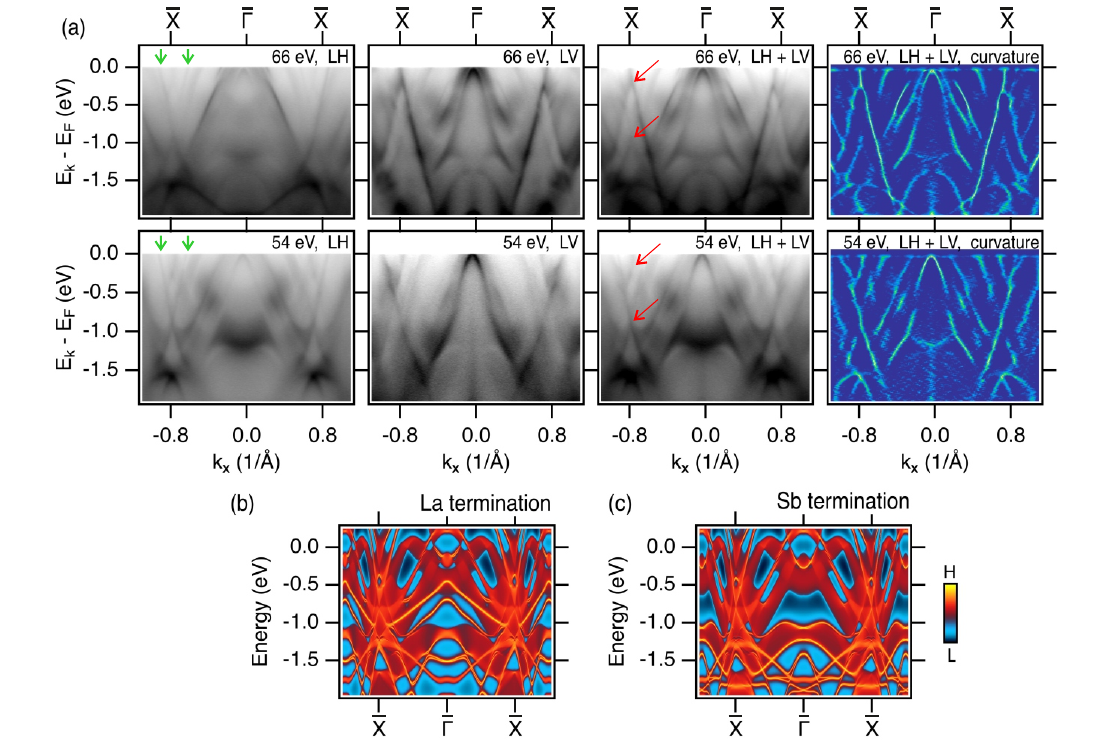}%
\caption{
Electronic structure along the $\bar{\text{X}}$--$\bar\Gamma$--$\bar{\text{X}}$ direction. (a) ARPES spectra collected with use of the 66 and 54 eV, LH and LV light polarizations. 
The curvature~\cite{45zhang2011} was calculated for the sum of both polarizations. 
Calculated surface spectrum for La (b) and Sb (c) surface terminations, obtained from surface Green function technique. 
In panel (a), green arrows show positions of the linear bands, while red arrows indicate Dirac-like crosses of the bands.
}
\label{f4}
\end{figure*}

\begin{figure*}
\centering
\includegraphics[width=\textwidth]{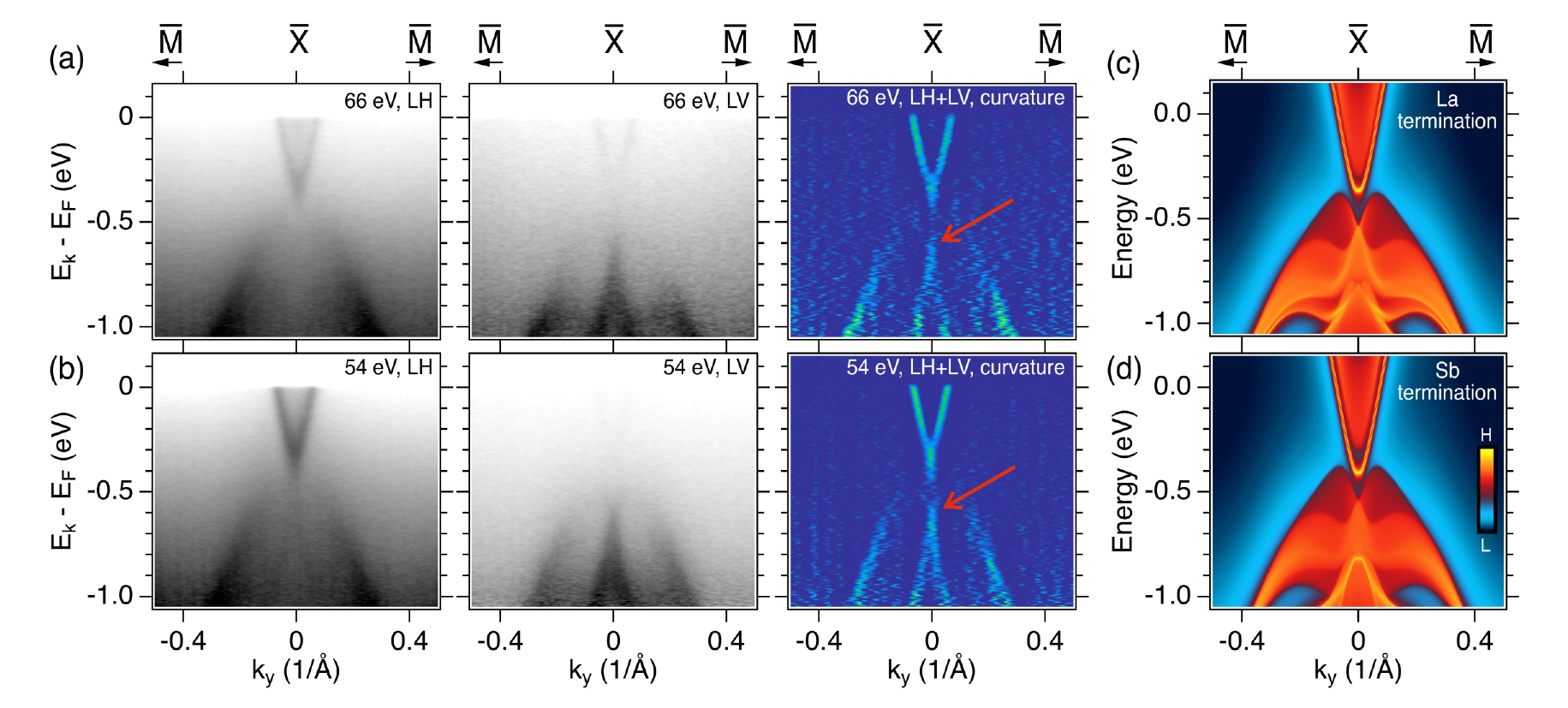}%
\caption{
The band structure along the $\bar{\text{M}}$--$\bar{\text{X}}$--$\bar{\text{M}}$ direction. 
(a) Measurements were made with a photon energy of 66 eV, horizontal and vertical polarization, respectively, and curvature was calculated for the sum of both polarizations. 
(b) Analogous to (a) for the energy of 54 eV. 
Red arrows indicated localization of Dirac points.
Surface Green function for (c) La and (d) Sb terminations along the $\bar{\text{M}}$--$\bar{\text{X}}$--$\bar{\text{M}}$ direction. 
Surface Green function for (e) La and (f) Sb termination along $\bar{\Gamma}$--$\bar{\text{X}}$--$\bar{\text{M}}$ direction. }
\label{f5} 
\end{figure*}

In Figure~\ref{f4} we present the dispersions of LaCuSb$_{2}$ along the $\bar{\text{X}}$--$\bar\Gamma$--$\bar{\text{X}}$ direction. 
Linear bands that cross at $\bar{\text{X}}$ point are clearly visible. 
With our high-quality data, we can clearly and convincingly observe the low-intensity linear bands previously reported for ARPES by Chamorro {\it et al.} in Ref.~\cite{35Chamorro2019} in this compound (green arrows in Figure~\ref{f4}a).
Moreover, our data reveal previously unseen details of the band structure, in particular the characteristic crossing of the linear bands forming an X shape. 
The intersection points of these linear bands are observed for binding energies of about $0.1$~eV and about $1$~eV, they were marked with red arrows in Fig.~\ref{f4}(a). 
These linear bands indicate the presence of Dirac fermions in the studied system, which is consistent with linear magnetoresistance and very low effective mass obtained in Shubnikov--de Haas experiment~\cite{30Ruszaa2018,35Chamorro2019}. 
Moreover, comparing the location of these points of intersection with bulk calculations [Fig.~\ref{f1}(d)], it can be presumed that they are part of the nodal line formed in the ${\text{X}}$--$\text{R}$ direction. 
The question of the existence of these nodal lines is continued further in the description of the structure measured in the direction $\bar{\text{M}}$--$\bar{\text{X}}$--$\bar{\text{M}}$. 
The observed electronic structure is very well reproduced by surface Green function calculations [Fig.~\ref{f4}(b) and (c)], except for the hole pocket appearing in calculations around the $\bar\Gamma$ point.
It is not visible in the experiment for the used photon energy and polarizations. 
The calculations were performed for two surface terminations [see Fig.~\ref{f4}(b) for La termination,  and Fig.~\ref{f4}(c) for Sb termination). 
Based on a few details (e.g. ``wavy'' band at $-1$~eV energy at $\bar\Gamma$ and parabolical band visible at $66$~eV with maxima at about $-1.5$~eV) it can be concluded that the measured spectra correspond to Sb termination. 

\begin{figure}
\includegraphics[scale=.65]{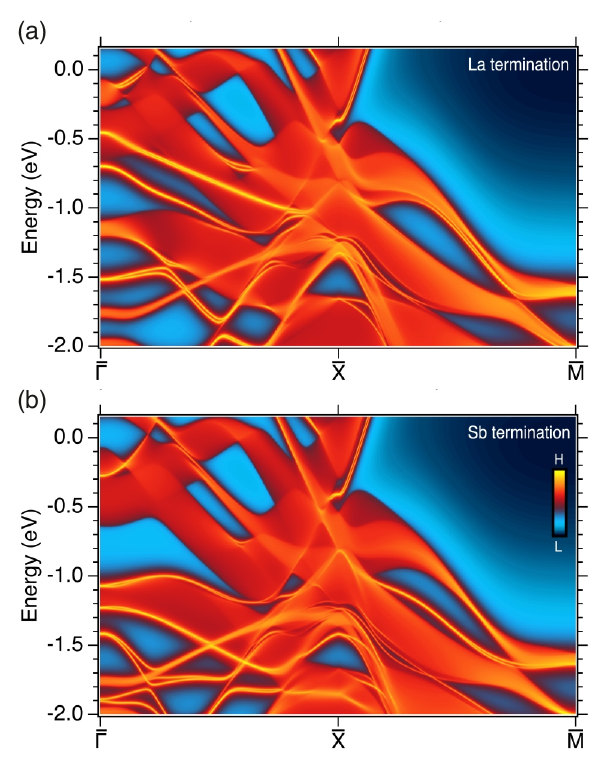}%
\centering
\caption{
Calculated surface spectrum for (a) La and (b) Sb terminations. 
}
\label{f6} 
\end{figure}

The dispersion along the $\bar{\text{M}}$--$\bar{\text{X}}$--$\bar{\text{M}}$ direction is shown in Figure~\ref{f5}. 
The ARPES results revealed a structure consisting of an electron pocket and other linear bands. The visible dispersion agrees well with the theoretical result shown in Fig.~\ref{f5}(c) and~\ref{f5}(d). 
Calculations suggest the existence of an energy gap between the upper electron like band and the band crossing both related to the nodal lines along the ${\text{X}}$--${\text{R}}$ direction. 
In fact, there is a lack of intensity in the measured spectra that matches the predicted gap, and it occurs at a binding energy of about $0.5$~eV. 
However, the quality of the data does not allow us to decide whether this is actually an energy gap or an effect of matrix elements. 
On the other hand, the linear crossing bands forming the Dirac point and consequently the nodal line are visible. 
The location of the Dirac points is indicated by red arrows in the curvatures in Fig.~\ref{f5}. 
The visibility of these bands strongly depends on the light polarization; the electron pocket crossing $E_{F}$ is clearly visible in LH, while the linear part of the structure is visible in LV. 
The bands measured at $66$~eV and $54$~eV are very similar, which proves their two-dimensional nature. 
To better understand the shape of the bands and the compatibility between the spectra collected along the $\bar{\text{X}}$--$\bar\Gamma$--$\bar{\text{X}}$ and $\bar{\text{M}}$--$\bar{\text{X}}$--$\bar{\text{M}}$ directions we presented the calculations along the $\bar{\Gamma}$--$\bar{\text{X}}$--$\bar{\text{M}}$ path in Figure~\ref{f6}. 
The lack of intensity around $0.5$~eV at $\bar{\text{X}}$ point fits the spectra collected in both directions. While the bands form the electron pocket in the $\bar{\text{M}}$--$\bar{\text{X}}$--$\bar{\text{M}}$ (Figure~\ref{f5}a) direction, they do not show an analogous dispersion in the $\bar{\text{X}}$--$\bar{\Gamma}$--$\bar{\text{X}}$ (Figure~\ref{f4}a) direction. 
In Figure~\ref{f4}a, this pocket corresponds to the increased intensity observed at $\bar{\text{X}}$ point in the energy range $-0.4$ to $0$~eV. 

\begin{figure*}
\includegraphics[width=\textwidth]{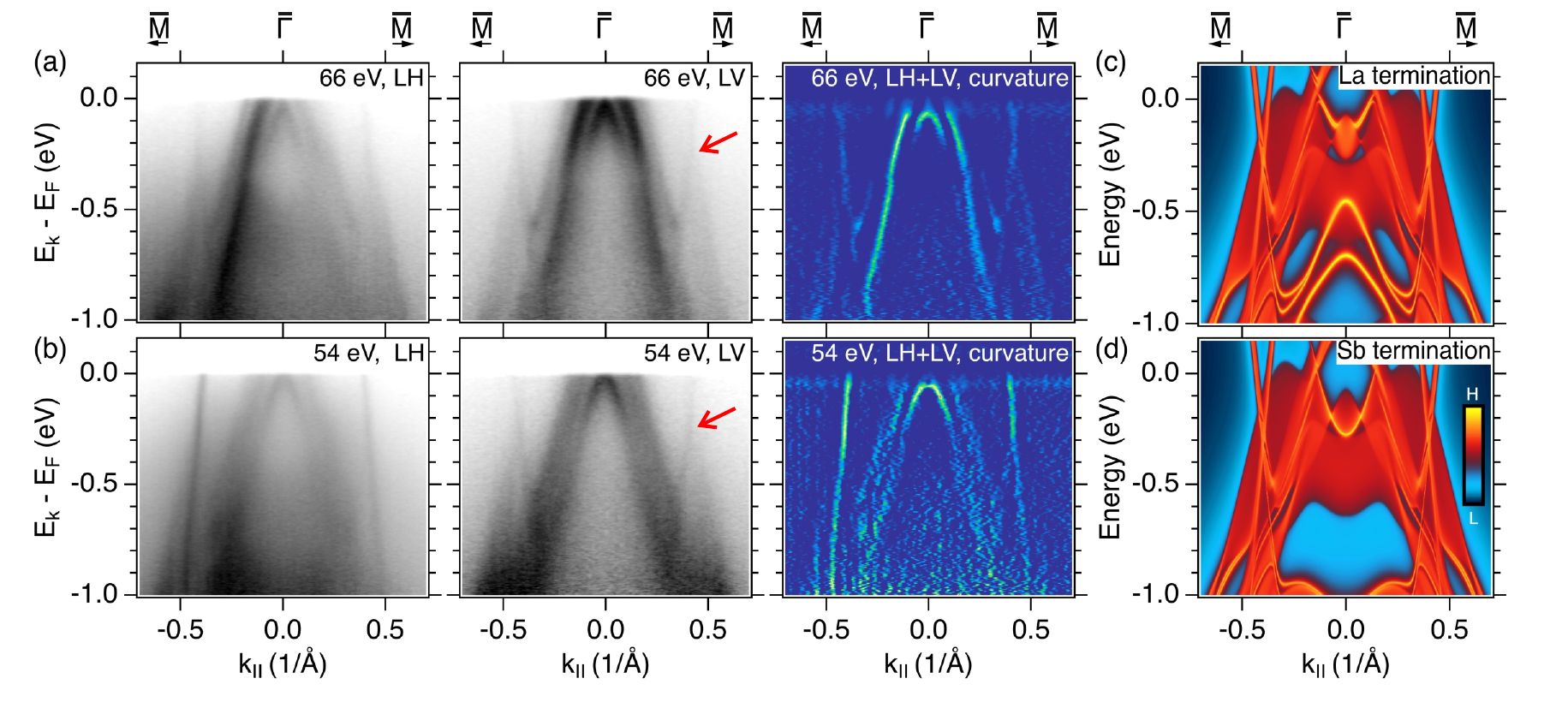}%
\caption{
The electronic structure along the $\bar{\text{M}}$--$\bar\Gamma$--$\bar{\text{M}}$ direction. 
Measurements were made with a photon energy of $66$~eV, horizontal and vertical polarization, respectively, and curvature was calculated for the sum of both polarizations. 
(b) Analogous to (a) for energy of $54$~eV. 
The red arrows show the location of the crosses of linear bands.
}
\label{f7} 
\end{figure*}
In the direction of $\bar{\text{M}}$--$\bar{\Gamma}$--$\bar{\text{M}}$, almost linear bands are visible crossing the Fermi level [Fig.~\ref{f7}(a) and (b)]. These bands cross and create a characteristic Dirac-like shape. 
Band-crossing is observed at about $0.2$~eV below the E$_{F}$ [marked with red arrows in Fig.~\ref{f7}(a) and (b)]. 
An analogous shape is visible in the surface Green function [Fig.~\ref{f7}(c) and (d)]. 
ARPES measurements indicate the gapless nature of this feature. 
According to the bulk calculation [Fig.~\ref{f1}(d)], the gap should be present, however in the surface Green function no gap is visible. 
As the surface function is an integration along the k$_{z}$ direction, it may mean that the hypothetical energy gap changes its location in the energy scale for different k$_{z}$ values. 
ARPES spectra for low excitation energies are sensitive to the surface and have a relatively low resolution in k$_{z}$, therefore they are better modeled by the surface Green function, which explains the lack of the energy gap.

\section{Summary}
\label{sec.sum}

In conclusion, we performed a detailed ARPES study supported by the theoretical calculation of LaCuSb$_{2}$ superconductor. FS consists of four branches, the ones around the $\bar{\Gamma}$ point are more 3-dimensional than the large sheets spanned between  $\bar{\text{X}}$ points and the agreement between the experiment and the theory is reasonable. 
We found Dirac like crossings at the $\bar{\text{X}}$ point 
corresponding to a nodal line along ${\text{X}}$--${\text{R}}$
as well as other linear dispersions along the $\bar{\text{M}}$--$\bar{\Gamma}$--$\bar{\text{M}}$ and $\bar{\text{X}}$--$\bar{\Gamma}$--$\bar{\text{X}}$ directions. The presence of Dirac fermions is in line with linear magnetoresistance and very low effective mass obtained in Shubnikov--de Haas experiment previously~\cite{35Chamorro2019}. Calculated surface Green function well represents details of the spectra indicating that Sb termination is studied. The experimentally observed nesting of the external FS branches appears to be stronger in LaCuSb$_{2}$ as compared to LaAgSb$_{2}$, which was not expected as CDWs appear in the latter system. The nesting found for the internal FS branches in LaAgSb$_{2}$ has a much larger wave vector than this corresponding to the CDW modulation.

\section{Acknowledgments}
P.P. and A.P. acknowledges the support by National Science Centre (NCN, Poland) under Project No.~2021/43/B/ST3/02166. 
This publication was partially developed under the provision of the Polish Ministry and Higher Education project ``Support for research and development with the use of research infra-structure of the National Synchrotron Radiation Centre SOLARIS'' under contract nr 1/SOL/2021/2. 
We acknowledge SOLARIS Centre for the access to the URANOS beamline, where the measurements were performed.



 \bibliographystyle{elsarticle-num-names} 
 \bibliography{Cu}





\end{document}